\documentstyle[epsfig]{aipproc}
\begin{document}
\title{Unique Quantum Stress Fields}

\author{Christopher L. Rogers and Andrew M. Rappe}
\address{Department of Chemistry and Laboratory for Research 
on the Structure of Matter, \\
University of Pennsylvania, Philadelphia, PA 19104-6323.}

\maketitle

\begin{abstract}
We have recently developed a geometric formulation of the stress field
for an interacting quantum system within the local density approximation
(LDA) of density functional theory (DFT). We obtain a stress field which is invariant with
respect to choice of energy density. In this paper, we explicitly
demonstrate this uniqueness by deriving the stress field for different
energy densities. We also explain why particular energy densities give 
expressions for the stress field that are more tractable than others, thereby
lending themselves more easily to first-principles calculations.
\end{abstract}

Understanding a material's energetic response to strain is 
fundamentally coupled to understanding the physics of a broad range of phenomena,
from surface reconstructions to piezoelectricity.
For example, it has been demonstrated that knowledge of the spatial 
distribution of microscopic stress via first-principles calculations can help explain
the onset of macroscopic polarization in a material \cite{ramer}. Therefore it is important
to understand the nature of stress at the microscopic level.

Formally, the electronic contribution to the microscopic stress
must be included via quantum mechanics. There have been
numerous methods developed for computing the quantum stress. (See
Ref. \cite{nielsen} and references contained within.) The stress field,
$\sigma_{\alpha \beta}(\mathbf{x})$, is a rank-two tensor field
usually taken to be symmetric (torque-free). The divergence of
$\sigma_{\alpha \beta}(\mathbf{x})$ must equal the force field
$F^{\alpha}(\mathbf{x})$ of the system:
\begin{equation}  
F^{\alpha}= \nabla_{\beta} \sigma^{\alpha \beta}. \label{force-stress}
\end{equation}
(Note that the Einstein summation convention for repeated indices is
used throughout the paper.) It is well-known that Eq.\
\ref{force-stress} does not uniquely define the stress field
(regardless of whether a system is quantum-mechanical or classical) 
since one can add any tensor whose divergence is
zero to $\sigma_{\alpha \beta}(\mathbf{x})$ and still recover the same
force field \cite{landau}. 

The volume-averaged or total stress $T_{\alpha\beta}$
has been related to the energetic response of the system to uniform
strain and to the integral over the stress field:
\begin{eqnarray}
T^{\alpha \beta} &=& \frac{\partial E}{\partial e_{\alpha \beta}} \\
                 &=& \int \sigma^{\alpha \beta}({\mathbf{x}}) d^3x, \label{FF}
\end{eqnarray}
where $e_{\alpha \beta}$ is a uniform scaling (strain) of the system.
Nielsen and Martin showed that $T_{\alpha \beta}$ is a unique well-defined
quantity which can be computed efficiently \cite{nielsen}. 
However Eq.\ \ref{FF} does not unambiguously define the stress field.
One can write the total energy $E$ as an integral over an energy density
$\mathcal{E}(\mathbf{x})$. Therefore
\begin{eqnarray}
T^{\alpha \beta} &=& \int d^3x \frac{\partial \mathcal{E}({\mathbf{x}})}{\partial e_{\alpha \beta}}.
\end{eqnarray}
Comparing Eq. 4 and Eq. 3 could lead to a definition of the stress field as \cite{FF,chetty} 
\begin{eqnarray}
\sigma^{\alpha \beta}({\mathbf{x}}) = \frac{\partial \mathcal{E}({\mathbf{x}})}{\partial e_{\alpha \beta}}. \label{FF2}
\end{eqnarray}
However, the total energy can also be expressed many other ways:
\begin{eqnarray}
\int d^3x \ {\mathcal{E}} + \int d^3x \nabla_{\alpha}W^{\alpha}.
\end{eqnarray}
If variations of $W^{\alpha}$ with respect
to the dynamical variables (e.g. single-particle wavefunctions) vanish
on the boundary of the system, then the minimum of $E$ remains invariant
with respect to choice of energy density. 
However, the energy density is changed by the presence of $W$, making
the expression of Eq.\ \ref{FF2} and Ref. 4 dependent on the choice
of energy density. Choosing a form of the energy density
is referred to as specifying an energy density ``gauge'', 
and hence the stress field of Eq.\ \ref{FF2} is said to be not gauge invariant.

We have recently developed a formulation of the quantum stress field which is invariant
to choice of energy density, and have used it to successfully calculate the stress fields 
in periodic systems using LDA-DFT \cite{rogers}. 
We define the stress field as 
\begin{eqnarray}
\sigma^{\alpha \beta}({\mathbf{x}}) = \frac{\delta E}{\delta \epsilon_{\alpha \beta}({\mathbf{x}})},
\end{eqnarray}
where $\epsilon_{\alpha \beta}({\mathbf{x}})$ is the strain tensor field.
This definition can be derived via a virtual work theorem \cite{landau}.
We then use a relationship known in continuum mechanics that equates $\epsilon_{\alpha \beta}$
with the Riemannian metric tensor $g_{\alpha \beta}({\mathbf{x}})$ \cite{fung,mistura}.
Therefore the stress field is
\begin{eqnarray}
\sigma_{\alpha\beta}&=&-\frac{2}{\sqrt{g}} \frac{ \delta E}{\delta g^{\alpha\beta}}, \label{sfield}
\end{eqnarray}
where ${g}$ and $g^{\alpha \beta}$ are the determinant and inverse 
of $g_{\alpha \beta}$, respectively.
In order to calculate $\sigma_{\alpha \beta}$, we first express the
constrained energy functional for LDA-DFT \cite{HK,KS} in a covariant manner via the principle of minimal coupling,
and then compute the functional derivative using a procedure well-known in general relativity
theory \cite{rogers,landau2}. It should be mentioned that this prescription for obtaining
$\sigma_{\alpha \beta}$ is not limited to LDA-DFT but can be generalized to derive the stress field
for other DFT functionals as well as explicit many-body systems
of quantum particles.


We now demonstrate the invariance of our expression for the stress
field with respect to choice of energy density.
In DFT there are two terms in the total energy expression 
which are commonly subjected to energy density gauge transformations.
The first is the single-particle kinetic energy $E_{\mathrm{k}}$ which can be written as
an integral over the so-called asymmetric kinetic energy density:
\begin{eqnarray}
E_{\mathrm{k}} &=& \int -\frac{1}{2} \sum_{i} 
\phi^{\ast}_i \partial_{\alpha} \left(\sqrt{g} g^{\alpha \beta} \partial_{\beta} \phi_i \right) d^3x, \label{asymkin}
\end{eqnarray} 
where the $\{ \phi_{i} \}$ are the single-particle wavefunctions which obey the
Kohn-Sham equations generalized for curvilinear coordinates \cite{rogers}. 
Therefore we are considering the total energy to be at a minimum. We consider here for brevity insulator
occupation numbers.
The variation of Eq.\ \ref{asymkin} can be written as
\begin{eqnarray}
\delta E_{\mathrm{k}} &=& \delta \left\{ \frac{1}{2} \int \sum_i 
\sqrt{g} \partial_{\alpha} \phi^{\ast}_{i} g^{\alpha \beta} \partial_{\beta} \phi_{i} d^3x \right \}
- \delta \left \{ \frac{1}{2} \int \sum_i 
 \partial_{\alpha} \left(\sqrt{g} \phi^{\ast}_{i} g^{\alpha \beta} \partial_{\beta} \phi_{i} \right)
d^3x \right \} \nonumber \\
&=&\delta \left\{ \frac{1}{2} \int \sum_i 
\sqrt{g} \partial_{\alpha} \phi^{\ast}_{i} g^{\alpha \beta} \partial_{\beta} \phi_{i} d^3x \right \}
- \delta \left \{ \frac{1}{2} \oint \sum_i 
 \sqrt{g} \phi^{\ast}_{i} g^{\alpha \beta} \partial_{\beta} \phi_{i} dS_{\alpha} \right \}, \label{kin}
\end{eqnarray}
where we have used the covariant version of the divergence theorem in the second line.
All variations in $\{ \phi_{i} \}$ vanish on the boundary of the system, and we require 
$\delta g^{\alpha \beta}=0$ on the boundary.
Therefore the surface term in Eq.\ \ref{kin} is zero. As a result the functional derivative required
for evaluating Eq.\ \ref{sfield} is
\begin{eqnarray}
\frac{\delta E_{\mathrm{k}}}{\delta g^{\alpha \beta}({\mathbf{y}})} &=& \frac{1}{2} 
\sqrt{g} \sum_{i}\partial_{\alpha} \phi^{\ast}_{i} \partial_{\beta} \phi_{i} + 
\frac{\partial \sqrt{g}}{\partial g^{\alpha \beta}} 
\left(\frac{1}{2}\sum_{i} \partial_{\alpha} \phi^{\ast}_{i} g^{\alpha \beta} \partial_{\beta} \phi_{i} \right) \nonumber \\
& & \mbox{} + \frac{1}{2} \sum_{i} \int d^3x \sqrt{g} \partial_{\gamma} \phi^{\ast}_{i} g^{\gamma \kappa}
\frac{\delta \left(\partial_{\kappa} \phi_{i} \right ) }{ \delta g^{\alpha \beta} ({\mathbf{y}})} \nonumber \\
& & \mbox{} + \frac{1}{2} \sum_{i} \int d^3x \sqrt{g} \partial_{\kappa} \phi_{i} g^{\gamma \kappa}
\frac{\delta \left(\partial_{\gamma} \phi^{\ast}_{i} \right ) }{ \delta g^{\alpha \beta} ({\mathbf{y}})}.
\end{eqnarray}
Note that this result is identical to what we would
obtain if we wrote $E_{\mathrm{k}}$ as an integral over the symmetric kinetic energy density:
\begin{eqnarray}
E_{\mathrm{k}} &=& \frac{1}{2} \sum_{i} \int \sqrt{g} \partial_{\alpha} \phi^{\ast}_{i} g^{\alpha \beta} 
\partial_{\beta} \phi_{i} d^3x,
\end{eqnarray}
and took the variation of this expression directly.
Therefore the stress field is invariant with respect to choice of kinetic energy density.
We mention that terms involving the variation of the wavefunctions with respect to 
metric vanish when the variation of the total ground-state energy is considered, since the
wavefunctions obey the Kohn-Sham equations \cite{rogers}.
In other words, requiring $\delta E/ \delta \phi^{\ast}_{i}=0$ implies
\begin{eqnarray}
\frac{\delta E}{\delta g^{\alpha\beta}({\mathbf{y}})} &=& \frac{\partial E}{\partial g^{\alpha\beta}({\mathbf{y}})} 
+ \sum_i \int d^3x \frac{\delta E}{ \delta\phi^{\ast}_i} \frac{ \delta\phi^{\ast}_{i}}{\delta g^{\alpha\beta}({\mathbf{y}})}
+\sum_i \int d^3x \frac{\delta E}{\delta \phi_i} \frac{ \delta \phi_i}{\delta g^{\alpha\beta}({\mathbf{y}}) } \nonumber \\
&=& \frac{\partial E}{\partial g^{\alpha\beta}({\mathbf{y}})}.
\end{eqnarray}

The Coulomb term describing the electrostatic electron-electron,
electron-ion, and ion-ion interactions in the total energy functional is also commonly 
subjected to energy density gauge transformations.
(This is entirely separate from transformations of the $U(1)$ electromagnetic gauge.)
We can define $E_{\mathrm{Coulomb}}$ as
\begin{eqnarray}
E_{\mathrm{Coulomb}} &=& \frac{1}{2}\int \sqrt{g} \rho V d^3x,
\end{eqnarray}
with
\begin{eqnarray}
\rho(\bbox{\mathrm{x}}) = \sum_i 
\frac{Z_i}{\sqrt{g}}\delta(\bbox{\mathrm{x}} - \bbox{\mathrm{R}}_i) -
n(\bbox{\mathrm{x}}),
\end{eqnarray}
where $Z_i$ is the charge of the $i$-th ion located at position
$\bbox{\mathrm{R}}_i$, and $n$ is the electronic charge density equal to
$\sum_i\phi^{\ast}_{i} \phi_{i}$. 
The potential $V$ can be computed from $\rho$ via the Poisson equation:
\begin{equation}
\frac{1}{\sqrt{g}} \partial_{\alpha}\left(\sqrt{g} g^{\alpha \beta}
\partial_{\beta}V \right) = -4\pi \rho. \label{Poisson}
\end{equation}
The variation of $E_{\mathrm{Coulomb}}$ is
\begin{eqnarray}
\delta E_{\mathrm{Coulomb}} &=& \delta \left \{ - \frac{1}{8 \pi} 
\int \partial_{\alpha}\left(\sqrt{g} g^{\alpha \beta}\partial_{\beta}V \right) V d^3x \right \} \nonumber \\
&=& \delta \left \{ \frac{1}{8 \pi} \int \sqrt{g} \ \partial_{\alpha}V g^{\alpha \beta} \partial_{\beta}V d^3x \right \}
- \delta \left \{ \frac{1}{8 \pi} \int \partial_{\alpha} \left( V \sqrt{g} g^{\alpha \beta}\partial_{\beta}V \right) d^3x 
\right \} \nonumber \\
&=& \delta \left \{ \frac{1}{8 \pi} \int \sqrt{g} \ \partial_{\alpha}V g^{\alpha \beta} \partial_{\beta}V d^3x \right \}
- \delta \left \{ \frac{1}{8 \pi} \oint  V \sqrt{g} g^{\alpha \beta}\partial_{\beta}V  dS_{\alpha}  \right \} \label{coul}.
\end{eqnarray}
Since the variations of the potential and the metric
vanish on the boundary, the surface term in Eq.\ \ref{coul} is zero. Therefore
the functional derivative of $E_{\mathrm{Coulomb}}$  with respect to metric is
\begin{eqnarray}
\frac{\delta E_{\mathrm{Coulomb}}}{\delta g^{\alpha \beta}({\mathbf{y}})} &=& 
\frac{1}{8 \pi} \sqrt{g} {\mathcal{F}}_{\alpha}{\mathcal{F}}_{\beta} + \frac{1}{8 \pi}
\frac{\partial \sqrt{g}}{\partial g^{\alpha \beta}} {\mathcal{F}}^{\gamma}{\mathcal{F}}_{\gamma}  \nonumber \\ 
&& \mbox{} + \frac{1}{4 \pi} \int \sqrt{g} g^{\gamma \kappa} {\mathcal{F}}_{\kappa} 
\frac{\delta {\mathcal{F}}_{\gamma}}{\delta g^{\alpha \beta}({\mathbf{y}})} d^3x, \label{coul2}
\end{eqnarray}
where the electric field ${\mathcal{F}}_{\alpha} = -\partial_{\alpha} V$. We would obtain
this same result if we initially expressed $E_{\mathrm{Coulomb}}$ as an integral over the Maxwell energy density:
\begin{eqnarray}
E_{\mathrm{Coulomb}} = \frac{1}{8 \pi} \int \sqrt{g} {\mathcal{F}}_{\alpha} g^{\alpha \beta}
{\mathcal{F}}_{\beta} d^3x.
\end{eqnarray}
This demonstrates that the stress field is invariant with respect to choice of
electrostatic energy density. 

The non-local term involving the variation of the electric field with respect metric
in Eq.\ \ref{coul2} is unwieldy and is in general difficult to compute. 
To eliminate this term it is advantageous to choose an energy density that
is the Lagrangian density in electromagnetism. Therefore
we write $E_{\mathrm{Coulomb}}$ as
\begin{eqnarray}
E_{\mathrm{Coulomb}} &=& \int \sqrt{g} \left ( \rho V - 
\frac{1}{8 \pi} {\mathcal{F}}^{\gamma}{\mathcal{F}}_{\gamma} \right) d^3x. \label{EML}
\end{eqnarray}
Now when $E_{\mathrm{Coulomb}}$ is varied, the term containing $\delta V$
will vanish due to the Poisson equation. 
Variation of Eq.\ \ref{EML} also gives a term relating to the variation of $\rho$ with respect to metric:
\begin{eqnarray}
\int \sqrt{g} \ V \frac{\delta \rho}{\delta g^{\alpha \beta}({\mathbf{y}})} d^3x. \label{rho}
\end{eqnarray}
As explained in the kinetic energy section, this variation (and those
of other energy terms) will be multiplied by $\delta E/\delta\phi^{\ast}_{i}({\mathbf{x}})$.
The Kohn-Sham equation insures that variations of $E$ with respect to
$\phi^{\ast}_{i}$ vanish, and the ionic charges are fixed, 
thereby removing the need to evaluate Eq.\ \ref{rho}.

In conclusion, we have demonstrated explicitly that our formulation for the quantum stress field within
DFT is invariant with respect to choice of energy density. Therefore the stress field
is a well-defined object that can be computed via first-principles to help gain a
microscopic understanding of stress-mediated phenomena in complex materials.

The authors wish to acknowledge R.\ M.\ Martin and D.\ H.\ Vanderbilt for
helpful discussions. This work was supported by
the Office of Naval Research under grant number N-00014-00-1-0372 
and the Air Force Office of Scientific Research, Air Force Materiel
Command, USAF, under grant number F49620-00-1-0170.


\begin{references}
\bibitem{ramer} N.J. Ramer, E.J. Mele, and A.M. Rappe, Ferroelectrics {\bf 206-207}, 31 (1998).
\bibitem{nielsen} O.H. Nielsen, and R.M. Martin, Phys. Rev. B. {\bf 32}, 3780 (1985).
\bibitem{landau} L.D. Landau and E.M. Lifshitz, \textit{Theory of Elasticity}, 3rd ed. (Butterworth-Heinemann, Oxford, 1986), pp. 1-7.
\bibitem{FF} A. Filippetti and V. Fiorentini, Phys. Rev. B. {\bf 61}, 8433 (2000).
\bibitem{chetty} N. Chetty and R.M. Martin, Phys. Rev. B. {\bf 45}, 6074 (1992).
\bibitem{rogers} C.L. Rogers and A.M. Rappe, Phys. Rev. Lett. (submitted 2000) cond-mat/0006274.
\bibitem{fung} Y.C. Fung, \textit{Foundations of Solid Mechanics}, (Prentice-Hall, Englewood Cliffs, 1965), pp. 90-92.
\bibitem{mistura} L. Mistura, J. Chem. Phys. {\bf 83}, 3633 (1985); Inter. J. Thermophys. {\bf 8}, 397 (1987).
\bibitem{HK} P. Hohenberg and W. Kohn, Phys. Rev. \textbf{136}, B864 (1964).
\bibitem{KS} W. Kohn and L. J. Sham, Phys. Rev. \textbf{140}, A1133 (1965).
\bibitem{landau2} L.D. Landau and E.M. Lifshitz, \textit{The Classical Theory of Fields}, 4th ed. (Pergamon, Oxford, 1975), pp. 270-273. 
\end{references}
\end{document}